\documentclass[12pt,a4paper,one column]{article}
\usepackage[utf8]{inputenc}
\usepackage{amsmath}
\usepackage{amsfonts}
\usepackage{amssymb}
\usepackage{graphicx}
\usepackage{graphics}
\usepackage{color}
\title{De-pinning of contact line of droplets on rough surfaces }
\author{V Madhurima* and K Nilavarasi**\\
Department of Physics\\
School of Basic and Applied Sciences\\
Central University of Tamil Nadu\\
Thiruvarur - 610101\\
email: *madhurima@cutn.ac.in, **nilavarasikv@gmail.com \\
}
\date { }
\begin{document}
\maketitle
%\tableofcontents
\section*{Abstract}
The present study reports the formation of self-assembled droplet pattern on the PDMS polymer coated over grooved side of DVD under saturated vapours of alcohols.  Comparison of the results with breath figures formed over unconstrained side of DVD is made. Four different environments namely methanol, ethanol, 2-propanol and n-butanol are used for the analysis. It is observed that the pattern formation occurs with methanol and ethanol vapours and not with 2-propanol and n-butanol. The difference is pattern formation with different alcohols is attributed to the variation in chain length and the presence of hydrophobic groups in alcohols, as given by Traube's rule. The distortion of patterns over constrained surface is attributed to the depinning of contact lines.

\section{Introduction}
The study of dynamics of a droplet’s contact line on structured surfaces is of importance for its applications in self-cleaning \cite{bhusan}, anti-corrosion \cite{chen}, bio-sensing \cite{chen} and so on. The contact line dynamics is greatly influenced by the presence of surface heterogeneities \cite{gennes}. These heterogeneities acts as droplet pinning sites. Although the dynamics of the contact line for hydrophobic/super-hydrophobic surfaces were extensively studied, the contact line dynamics during the phase transition process like evaporation and condensation were relatively less studied \cite{chen, marin}. \\

Breath figure is one such process involving the phase change process of evaporation and condensation to form self-assembled droplet patterns \cite{bunz,stenzel,ma, escale, steyer, pitios}. Usually breath figures are formed with super saturated vapor of water. Ding et al., was the first to report the formation of self-assembled droplet pattern using non-aqueous vapors of ethanol and methanol \cite{ding}. They proved that the breath figure method is not only suitable for water vapor but is also compatible with organic solvents like ethanol and methanol. But the influence of constraints on the underlying surface on the formation of breath figures using non-aqueous vapor is so far not reported to the best of our knowledge.\\ 

Alcohols are the simplest form of amphiphilic molecules having hydrophobic and hydrophilic groups \cite{ballal}. These hydrophobic and hydrophilic group of the alcohols play an important role in deciding the contact area of the droplets. When a droplet of liquid is impinged on the surface (preferentially solid/soft solid surface), the hydrophobic group of an added alcohol preferentially goes to the surface \cite{ballal}. The hydrophobicity of the alcohols increases with increase in the chain length of alcohols and at the interface, surface tension decreases with increase in number of $CH_{2}$ group according to Traube’s rule \cite{ruth}.\\

The present paper deals with the experimental verification of the effect of hydrophobic interactions in the formation of self-assembled droplet formation and Traube’s rule of surface tension at interfaces. The effect of contact line dynamics of these alcohols on the constrained surface is also investigated. \\

\section{Theory}
When a liquid droplet is placed on a solid surface having heterogeneities, the contact line can stay pinned until the contact angle reaches a critical angle \cite{marmur}. The contact angle for smooth surface under thermodynamic equilibrium conditions is defined by Young’s equation, which is given as \cite{young},
\begin{equation}
	cos\theta_{Y}  =  \frac{\gamma_{sv}-\gamma_{sl}}{\gamma_{lv}}
	 \label{eq:}
\end{equation}
	 where  $\gamma_{sv}$, $\gamma_{sl}$ and $\gamma_{lv}$ are the solid-vapour, solid-liquid and liquid-vapor interfacial tensions respectively and $\theta_{Y}$ is the Young's contact angle  which is the intrinsic contact angle of the surface at the three phase contact line. For a heterogeneous rough surface, the contact angle can vary over a finite range \cite{kalinin, marmur}.  On a smooth surface, the contact line will not advance (recede) until the contact angle reaches a critical value of the advancing (receding) contact angle. The difference between the advancing and receding contact angle is called contact angle hysteresis, which characterizes the strength of pinning of contact line at the asperities of the rough surface \cite{kalinin}.\\

After pinning the droplets start to evaporate. There are two different modes of evaporation of a liquid droplet on a polymer surface, namely, the constant contact radius (CCR) mode and the constant contact angle (CCA). The CCR mode is characterized by the pinned contact line and decreased contact angle whereas the CCA mode is characterized by the receding contact line and constant contact angle \cite{chen}.\\

The behavior of droplet evaporation is also affected by the surface morphology and chemical composition.  Orejon et al.\cite{orejon} in their investigation of the dynamics of the three-phase contact line of evaporating droplets on different surfaces with varying substrate hydrophobicity proposed that the depinning of contact line on rough surfaces occurs when the unbalanced Young’s force is large enough to overcome an intrinsic energy barrier arising from surface roughness. CCR-CCA transition, a consequence of the pinning-depinning phenomenon was observed during the evaporation on the surfaces with relatively large stripe widths \cite{wang}. 

\section{Experimental details} 
\subsection{Materials and methods}
Commercially available digital versatile disk was used as substrate in the present study. In the present study, poly-dimethyl siloxane (PDMS) from Sigma-Aldrich and chloroform (purity 99.9 \%) purchased from Merck (Emplura) were used as polymer and solvent respectively. Methanol, ethanol, n-butanol and 2-propanol purchased from Merck (Emplura) were used as environment. \\
The aluminium layer in the DVD were peeled off using forceps and the side facing the aluminium layer was used as constrained surface consisting of parallel grooves. The other side was used as the smooth surface.\\
For the preparation of patterned films using the supersaturated vapor of methanol, 5 ml of methanol is added to the petri-dish and kept inside the glass chamber. The polymer solution was prepared and casted on the surface of the disk. The casted disk was then placed inside the chamber containing methanol vapors. After complete evaporation of solvent, the films were characterized using confocal laser scanning microscope (CLSM). Same procedure is followed to prepare patterned films with other vapors namely ethanol, 2-propanol and n-butanol.\\

\subsection{Mechanism of formation of self-assembled droplet patterns}
In the beginning of the formation of self-assembled droplet pattern, the non-aqueous vapor starts to condense due to evaporation of the solvent. The droplets condense on the polymer solution surface. Spreading and dissolving of droplets in the polymer solution shouldn’t take place if pores are to be formed. The spreading behavior of the droplet on the surface is defined using spreading co-efficient as, 
\begin{equation}
S =\gamma_{s}-(\gamma_{l}+\gamma_{ls})
\label{eq:}
\end{equation}
where $\gamma_{s}$ and $\gamma_{l}$ are the surface tension of the solution and  condensed liquid respectively. $\gamma_{ls}$ is the interfacial tension of the liquid and solution.  $S>0 $ indicates spreading behaviour is more and $S<0$ indicates non-spreading behaviour of the liquid. Once the solvent and liquid completely evaporates, the polymer film is left with array of pores. \\ 

\section{Results and Discussion}
\begin{figure}[h]
\centerline{\includegraphics[width=5in]{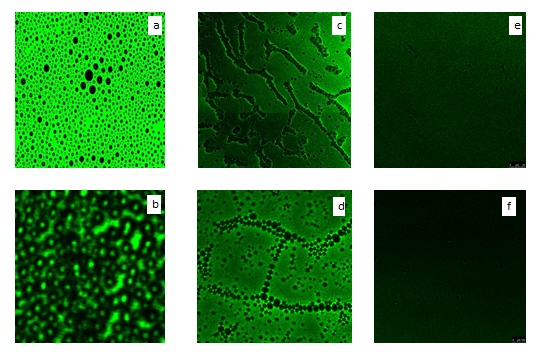}}
\caption[]{Self-assembled droplet patterns formed on a. smooth surface with ethanol vapor, b. smooth surface with methanol vapor, c. constrained surface with ethanol vapor d. constrained surface with methanol vapor, e. surfaces with propanol vapor and f. surfaces with butanol vapor.  }%
\label{1}
\end{figure}

The patterned films on smooth and constrained surfaces using methanol, ethanol, 2-propanol and n-butanol vapours were characterized for surface morphology using CLSM and are shown in Figure~\ref{1}. With methanol and ethanol vapours, the films were patterned with pores. The smooth surface showed a hexagonal array of self-assembled droplet pattern as expected due to Marangoni convection and formation of a lubricating thin polymer layer as a result of instant precipitation of the polymer\cite{mohan}. The instant precipitation prevents the droplets from coalescing with each other thus forming hexagonal pore patterns. The CLSM images of grooved surface showed a different pattern of pores. This distortion in pattern formation is ascribed to the depinning mechanism of contact lines. This depinning occurs due to the presence of constraints on the underlying surface \cite{vm, gennes}. \\
For a droplet on a rough surface having number of grooves parallel to each other, the contact line can align in two ways. It can lie either parallel to the grooves and get pinned or lies at an angle to the grooves. In the latter case, the contact line is displaced continuously without any pinning \cite{gennes}. This prevents the self-assembly of droplets into a hexagonal pattern. As a result, a distorted hexagonal ring of pores with cluster of pores inside is observed.\\
\begin{figure}[h]
\centerline{\includegraphics[width=5in]{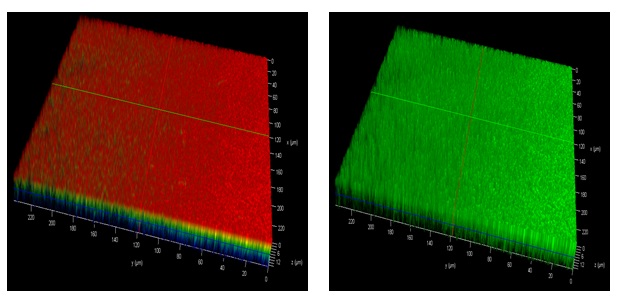}}
\caption[]{3D images of surfaces with propanol and n-butanol vapours.  Confirming the presence of polymer film over the substrate. }%
\label{2}
\end{figure}

The CLSM images of surfaces in 2-propanol and n-butanol environment doesn’t show any patterns and is shown in Figure~\ref{1}.  The 3D images of the polymer film formed over the substrate is shown in Figure~\ref{2}.  For 2-propanol and n-butanol, there is an increase in $CH_{2}$ group. This increase in $CH_{2}$ group reduces the surface tension which in turn affects the contact area of the droplet.  The increase in droplet contact area favours the coalescence of droplets. On complete evaporation of the solvent, a layer of polymer film without any significant patterns is observed. This result also proves the Traube’s rule relating surface tension of organic compounds to the number of hydrocarbon $CH_{2}$ groups present in the molecule. According to Traube’s rule, for every extra $CH_{2}$ group, the reduction in surface tension triples \cite{ruth}.    

\section{Conclusion}
The formation of self-assembled droplet pattern on the PDMS polymer coated over grooved side of DVD under saturated vapors of alcohols is reported. It is observed that with the increase in alkyl chain of alcohols, the formation of droplets pattern decreases.  Hexagonal pattern of self-assembled droplets on smooth surface was observed for surfaces patterned under methanol and ethanol vapors. A distorted hexagonal ring with cluster of pores inside was observed for grooved surface under same vapor conditions. The distortion in pattern formation is attributed to the depinning effect of contact line on rough surface.  With 2-proppanol and n-butanol, no patterns of droplets were observed.  This confirms the TRAUB's rule of decrease in surface tension with increase in $CH_{2}$ group.

\end{document}